# Chemical nitrogen fractionation in dense molecular clouds

*Jean-Christophe Loison[1]\*, Valentine Wakelam[2], Pierre Gratier[2], Kevin M. Hickson[1]*

\*Corresponding author: jean-christophe.loison@u-bordeaux.fr

[1] *Institut des Sciences Moléculaires (ISM), CNRS, Univ. Bordeaux, 351 cours de la Libération, 33400, Talence, France*
[2] *Laboratoire d'astrophysique de Bordeaux, Univ. Bordeaux, CNRS, B18N, allée Geoffroy Saint-Hilaire, 33615 Pessac, France.*

**ABSTRACT**

Nitrogen-bearing molecules display variable isotopic fractionation levels in different astronomical environments such as in the interstellar medium or in the Solar System. Models of interstellar chemistry are unable to induce nitrogen fraction in cold molecular clouds as exchange reactions for $^{15}N$ are mostly inefficient. Here, we developed a new gas-grain model for nitrogen fractionation including a thorough search for new nitrogen fractionation reactions and a realistic description of atom depletion onto interstellar dust particles. We show that, while dense molecular cloud gas-phase chemistry alone leads to very low fractionation, $^{14}N$ atoms are preferentially depleted from the gas-phase due to a mass dependent grain surface sticking rate for atomic nitrogen. However, assuming an elementary $^{14}N/^{15}N$ ratio of 441 (equal to the solar wind value), our model leads to only low $^{15}N$ enrichment for all N-containing species synthesized in the gas-phase with predicted $^{14}N/^{15}N$ ratios in the range 360-400. Higher enrichment levels can neither be explained by this mechanism, nor through chemistry, with two possible explanations. (I) The elementary $^{14}N/^{15}N$ ratio in the local ISM is smaller, as suggested by the recent work of Romano et al, with an hypothetic $^{15}NNH^+$ and $^{15}NNH^+$ depletion due to variation of the electronic recombination rate constant variation with the isotopes. (II) $N_2$ photodissociation leads to variable nitrogen fractionation in diffuse molecular clouds where photons play an important role, which is conserved during dense molecular cloud formation as suggested by the work of Furuya & Aikawa.

# 1 INTRODUCTION

Planetary systems, such as our own Solar System arise from the collapse of a dense molecular cloud (Füri & Marty 2015). Consequently, remnants of earlier evolutionary stages might be expected to be preserved in the composition of Solar System objects. Isotopic abundances are a sensitive probe of the processes leading to the formation of planetary systems as hydrogen, carbon, nitrogen and oxygen often present large differences in their fractionation ratios between the various steps. Among them, nitrogen fractionation is a key issue. The N/$^{15}$N ratio in the Solar system is equal to 441 ± 2.5 for the main reservoir of N (the Sun's photosphere (Marty et al. 2010)), a value which is seemingly incompatible with the elemental ratio in the local interstellar medium (ISM) estimated to be 290 ± 40 (Adande & Ziurys 2012) from interpolation of the galactic gradient using CN/C$^{15}$N and HNC/H$^{15}$NC measurements across the galaxy. This low elemental N/$^{15}$N ratio also seems to agree with the observed CN/C$^{15}$N and HCN/HC$^{15}$N ratios in diffuse molecular clouds (Ritchey et al. 2011, Lucas & Liszt 1998). Such a low elemental ratio may be due to galactic chemical evolution (Romano et al. 2017). Nevertheless, (Colzi et al. 2018) found recently a N/$^{15}$N ratio in HCN and HNC across the galaxy much closer to 400, which challenges the Adande and Ziurys (2012) result. Moreover, the measurements may be affected by efficient nitrogen fractionation processes in diffuse or/and dense molecular clouds. To test for possible chemical $^{15}$N fractionation in dense molecular clouds, Terzieva and Herbst (2000) developed a chemical model introducing various $^{15}$N exchange reactions. Their model resulted in low $^{15}$N enrichment, although further studies using their chemical network derived more notable fractionations (Rodgers & Charnley 2008a, Rodgers & Charnley 2008b, Wirström et al. 2012). Subsequent work by Roueff et al. (2015) however, clearly demonstrated that the efficient fractionation reactions proposed by Terzieva and Herbst (2000) are characterized by activation barriers and are therefore inefficient at low temperature. Furthermore, recent simulations by Wirström and Charnley (2018) showed that the additional fractionation reactions proposed by Roueff et al. suppressed $^{15}$N enrichment in molecules containing the nitrile functional group. To date, no astrochemical model based on gas-phase chemistry alone has been able to induce nitrogen fractionation levels greater than a few percent. Although the Wirström and Charnley (2018) model included simplified grain processes such as the depletion of gas-phase species onto grains; grain surface chemistry and grain desorption (with the exception of H$_2$ desorption) were not included. In addition, N and N$_2$ depletion was switched off, so any processes induced by nitrogen sticking on grains or by surface chemistry could not be reproduced.

In the absence of alternative viable fractionation reactions, an alternative pathway to consider is $N_2$ photodissociation as shown by Heays *et al.* (2014) where N-fractionation arises from to the more efficient photodissociation of $^{15}NN$ compared with $N_2$ due to the self-shielding of $N_2$ at high abundances. In dense molecular clouds however, photons are almost entirely generated by emission from electronically excited $H_2$, with fluxes that are too low for this type of process to be efficient. However, Furuya and Aikawa (2018) have developed a model to simulate molecular cloud formation starting from a diffuse H I cloud. In the first part of cloud evolution, UV radiation is high enough to ensure photodissociation. A key point in their model is the quick transformation of N atoms into $N_2$ and s-$NH_3$ (s- meaning species on grains) for low visual extinction levels ($A_v$ below 2). Then photodissociation of $N_2$ is important and self-shielding of $N_2$ leads to preferential $N^{15}N$ depletion and $^{15}N$ enrichment in the gas phase. The $N^{15}N$ depletion is partly conserved in the dense cloud with large $A_v$ but their model cannot reproduce the various observations. Indeed, the calculated $N_2H^+/^{15}NNH^+$ and $N_2H^+/N^{15}NH^+$ ratios are around 410, far smaller than the high $N_2H^+$ depletion levels observed in L1544 and L429 (Redaelli *et al.* 2018). In their model, nitrogen atoms are rapidly removed from the gas phase and transformed into s-$NH_3$ (on grains), which is mostly retained on the surface. Then, the gas phase nitrogen chemistry in dense molecular clouds is initiated by dissociation of $N_2$ through cosmic rays or by reaction with $He^+$. The Furuya & Aikawa model is indeed very interesting but involves significant approximations due to the importance of grain chemistry in their model, grain chemistry which is not well characterized at all. A better knowledge of grain chemistry, particularly for $A_v$ below 2 where photons can not only photo-desorb species but also induce chemical desorption through photodissociation and recombination, are clearly needed.

A critical point for the comparison between observed fractionation levels and astrochemical models are the observational uncertainties. Indeed, for some molecules, the main isotopologue transitions are optically thick, leading to a potential underestimation of abundances, or conversely, uncertainties can be large due to the very weak intensities of the minor isotopologue lines. For HCN, HNC, and CN, the extent of nitrogen fractionation is deduced almost always by the so called double isotopes ratio method deducing the $^{14}N/^{15}N$ ratio from $H^{13}CN/HC^{15}N$, $HN^{13}C/H^{15}NC$ and $^{13}CN/C^{15}N$ ratios using a $^{12}C/^{13}C$ ratio equal to 68 (Milam *et al.* 2005, Adande & Ziurys 2012). However, Roueff et al. (2015) showed that carbon chemistry leads to variable, and time dependent $^{13}C$ fractionation levels in dense molecular clouds, so that $^{15}N$ enrichment levels obtained using the double isotopes ratio method are unreliable. If we ignore the double isotope determinations and most of the cases where the main isotopologue emission lines are highly saturated (requiring complex analyses leading to very high uncertainties (Gonzalez-Alfonso & Cernicharo 1993, Daniel *et al.* 2013)), the range of observed N/$^{15}$N ratios in dense molecular clouds is much less variable (around 250-350 taking into account the uncertainties, see Table I), apart from the notable exceptions of N$_2$H$^+$ in L1544 and HCN and HNC in B1. The results of (Daniel et al. 2013) for B1 are doubtful as noted by the authors themselves. Indeed, even if they accounted for the opacity of the lines in a similar manner to (Magalhães *et al.* 2018) the lines of the main isotopologues were highly saturated. It should be noted that we do not present the results for NH$_2$D/$^{15}$NH$_2$D observations (see for exemple (Gerin *et al.* 2009)) as we do not consider deuterium chemistry in this work. Indeed, $^{15}$NH$_2$D formation reaction rate constants are not identical to those of NH$_2$D (Roueff et al. 2015).

**Table 1**: Observations for $^{15}$N in dense molecular clouds (we include the measurements where the main isotopologue emission lines show small opacity but the cases where the main isotopologue emission lines show very strong opacity nor the ones using the double isotopes methods are in most of the cases not reported here). The values given in brackets are errors reported in the literature which may be strongly underestimated.

| species | ratio | Cloud | references |
|---|---|---|---|
| N$_2$H$^+$/$^{15}$NNH$^+$ | $1000^{+260}_{-220}$ | L1544 | (Redaelli et al. 2018) |
| | > 600 | B1 | (Daniel et al. 2013) |
| | 320(60) | OMC-2 | (Kahane *et al.* 2018) |
| | $700^{+210}_{-140}$ | L694-2 | (Redaelli et al. 2018) |
| N$_2$H$^+$/N$^{15}$NH$^+$ | $920^{+300}_{-200}$ | L1544 | (Redaelli et al. 2018) |
| | $400^{+100}_{-65}$ | B1 | (Daniel et al. 2013) |
| | $330^{+170}_{-100}$ | L16293E | (Daniel *et al.* 2016) |
| | 240(50) | OMC-2 | (Kahane et al. 2018) |
| | $670^{+150}_{-230}$ | L183 | (Redaelli et al. 2018) |
| | 730(250) | L429 | (Redaelli et al. 2018) |
| | $580^{+140}_{-110}$ | L694-2 | (Redaelli et al. 2018) |
| NH$_3$/$^{15}$NH$_3$ | 330(50) | B1 | (Lis *et al.* 2010) |
| | $300^{+55}_{-40}$ | B1 | (Daniel et al. 2013) |
| CN/C$^{15}$N | $240^{+135}_{-65}$* | B1 | (Daniel et al. 2013) |
| | 270(60)* | OMC-2 | (Kahane et al. 2018) |
| HCN/HC$^{15}$N | $165^{+30}_{-25}$** | B1 | (Daniel et al. 2013) |
| | 338(28)* | L1498 | (Magalhães et al. 2018) |
| HNC/H$^{15}$NC | $75^{+25}_{-15}$** | B1 | (Daniel et al. 2013) |
| HNC/H$^{15}$NC | 250-330* | TMC1 | (Liszt & Ziurys 2012) |
| HC$_3$N/HC$_3^{15}$N | 257(54) | TMC1 | (Taniguchi & Saito 2017) |
| | 338(12) | L1527 | (Araki *et al.* 2016) |
| | 400(20) | L1544 | (Hily-Blant *et al.* 2018) |
| HC$_5$N/HC$_5^{15}$N | 344(53) | TMC1 | (Taniguchi & Saito 2017) |

*: main isotopologue shows opacity
**: main isotopologue shows very strong opacity

To test the influence of the updated network in the gas phase and the surface chemistry on nitrogen fractionation, we have developed a new gas-grain model including relevant chemical and physical processes. The chemical model including the various updates is presented in

Section 2 and the comparison with observations and previous models is described in Section 3. Our conclusions are presented in Section 4.

## 2 THE CHEMICAL MODEL

To compute the abundances we use a new chemical network based on the chemical model Nautilus in its 3-phase version (Ruaud *et al.* 2016) using kida.uva.2014 (Wakelam *et al.* 2015) with updates from Loison *et al.* (2016), Loison *et al.* (2017) and Vidal *et al.* (2017) for the chemistry. In the 3-phase model we make a distinction between the first 2 monolayers of molecules on top of the grains (considered as the surface) and the rest of the layers below (considered as the bulk), the gas being the third phase. In this model, only molecules at the surface can desorb (direct thermal desorption, chemical desorption, cosmic-ray induced desorption, and photodesorption). Species can however diffuse in both the surface and the bulk but with a much smaller efficiency in the bulk. To compute the surface probability of reaction for reactions with activation barriers, we consider the competition between reaction, diffusion and evaporation. Diffusion by tunneling is considered only for reactions involving hydrogen atoms. Cosmic-ray induced photodissociations can occur through both grain phases. All details of the equations and model parameters not specifically discussed in this article are the same as in Ruaud et al. (2016).

The new network is limited to a carbon skeleton up to $C_3H_xN_y$ (x = 0-2, y=0-1) and $C_3H_xN_y^+$ (x = 0-2, y=0-1), to reduce the number of reactions when considering all $^{15}N$ species. It includes 4641 reactions in the gas phase and 5154 reactions on grains including the reactions between the surface and the bulk as well as photodissociation. The new network reproduces the abundances of the complete network for the main species studied here. We have also introduced $^{15}N$ exchange reactions using an updated version of the network presented in Roueff et al. (2015) and some new exchange reactions using recent published works.

The surface network is similar to the one in Ruaud *et al.* (2015) with some updates from Wakelam *et al.* (2017). Following Hincelin *et al.* (2015), the encounter desorption mechanism is included in the code. This mechanism accounts for the fact that the $H_2$ binding energy on itself is much smaller than on water ices and prevents the formation of several $H_2$ monolayers on grain surfaces.

The accretion rate for a species *i* is described by the following classical expression

$$R_{acc}(i) = \sigma \sqrt{\frac{8kT}{\pi\mu}} n(i) n_d$$

where σ is the collision cross-section between the species *i* and the grain, √(8kT/πμ) is the relative average speed of the species *i* and the grain (*k* is the Boltzmann constant, μ is the reduced mass of the species *i* and the grain so mainly the mass of species *i*), n(*i*) is the density of the species *i* and $n_d$ is the grain density.

For desorption from the surface, we consider thermal desorption and desorption induced by cosmic-rays (Hasegawa & Herbst 1993, Dartois *et al.* 2015), as well as by exothermic chemical reactions (the exothermicity of surface chemical reactions allows for the species to be desorbed after their formation) (Garrod *et al.* 2007). The Garrod et al. (2007) chemical desorption mechanism leads to approximately 1% desorption of the newly formed species with 99% remaining on the grain surfaces (this corresponds to an *a* factor of 0.01 in Garrod et al. (2007)). The binding energies of species to the surface have been updated from Wakelam et al. (2017). These values are considered to be isotope independent as they are controlled through interactions which vary as a function of their electronic properties which are identical for isotopologues rather than as a function of the mass of the species.

The chemical composition of the gas-phase and the grain surfaces is computed as a function of time. The gas and dust temperatures are equal to 10 K, the total $H_2$ density is equal to $2\times10^4$ cm$^{-3}$ (various runs have been performed with the total H density varied between $2\times10^4$ cm$^{-3}$ and $2\times10^5$ cm$^{-3}$). The cosmic-ray ionization rate is equal to $1.3\times10^{-17}$ s$^{-1}$ and the total visual extinction ($A_v$) is set equal to 10. All elements are assumed to be initially in atomic form except for hydrogen, which is entirely molecular. Elements with an ionization potential below the maximum energy of ambient UV photons (13.6 eV, the ionization energy of H atoms) are initially in a singly ionized state, i.e., C, S and Fe. The initial abundances shown in Table 2 are similar to those of Table 1 of Hincelin *et al.*(2011), the C/O elemental ratio being equal to 0.7 in this study. In the nominal version of the model, sulphur is not depleted leading to better agreement with observations (Vidal et al. 2017, Fuente *et al.* 2016). However, a run was performed with a sulphur depletion factor of 10, allowing us to conclude that sulphur chemistry has a negligible effect on nitrogen fractionation. Dust grains are considered to be spherical with a 0.1μm radius, a 3 g.cm$^{-3}$ density and about $10^6$ surface sites, all chemically active. The dust to gas mass ratio is set to 0.01.

**Table 2.** Elemental abundance (/H).

| Element | Abundance |
|---------|-----------|
| He | 0.09 |
| C | 1.7e-4 |
| N ($^{14}$N) | 6.2e-5 |
| $^{15}$N | 1.406e-7 |
| $^{14}$N/$^{15}$N | 441 |
|  | 300 |
| O | 2.4e-4 |
| S | 1.5e-5 |
| Fe | 1.00e-8 |

2.1 $^{15}$N exchange reactions

In the interstellar medium, some fractionation may occur due to the fact that zero-point energy (ZPE) differences favor one direction (reverse or forward) for barrierless exchange reactions (such as $^{15}$N$^+$ + N$_2$ → N$^+$ + $^{15}$NN). The rate constants for the fractionation reactions have been studied in detail (Terzieva & Herbst 2000, Roueff et al. 2015). Using the work of Henchman and Paulson (1989), we consider in this work that all reactions involve adduct formation which correspond to reactions of type B in (Roueff et al. 2015). Then (f = forward = reaction toward the right in Table 3, r = reverse = reaction toward the left in Table 3):

$k_f = \alpha \times (T/300)^\beta \times f(B,m)/(f(B,m) + \exp(\Delta E/kT))$,

$k_r = \alpha \times (T/300)^\beta \times \exp(\Delta E/kT)/(f(B,m) + \exp(\Delta E/kT))$

with α and β given by capture theory for barrierless reactions, f(B,m) is related to the symmetry of the system (Terzieva & Herbst 2000) and ΔE = exothermicity of the reactions (see Table 3).

We checked that the $^{15}$N + HCNH$^+$ and $^{15}$N + N$_2$H$^+$ reactions have large barriers in the entrance valley as found by Roueff et al. (2015), so that these processes become negligibly small at low temperature. We searched for additional processes that could lead to nitrogen fractionation although no efficient ones were found. This is due to the fact that nitrogen is mainly present as neutral N atoms, which are reactive only with radicals (Dutuit *et al.* 2013) and a few ions; reactions having bimolecular exit channels leading to N$_2$, CN, HCN, HCN$^+$/HNC$^+$ and HCNH$^+$, which prevent efficient isotopic exchange. Among them, the HCNH$^+$ + HC$^{15}$N and NH$_4^+$ + $^{15}$NH$_3$ reactions favor nitrogen fractionation for HCNH$^+$ and NH$_4^+$. However, in dense molecular clouds, the abundances of HCN and NH$_3$ are too low for these reactions to lead to significant fractionation. For the HCNH$^+$ + H$^{15}$NC reaction, once the adduct is formed, the most exothermic exit channels will always be favored, that is, HC$^{15}$NH$^+$ + HCN and HCNH$^+$ + HC$^{15}$N at equilibrium (both equal to 0.5 as the energy corresponds to the exothermicity of the HNC → HCN isomerization) so that this reaction does not induce nitrogen fractionation.

Roueff et al. (2015) proposed that the N + CN reaction may play a role in nitrogen fractionation. Wirström and Charnley (2018) have shown than the inclusion of the $^{15}$N + CN → N + C$^{15}$N reaction leads to some $^{15}$N enrichment in nitriles. However, considering the bimolecular exit channel of the $^{15}$N + CN reaction (Daranlot *et al.* 2012), it is highly unlikely that this reaction leads to efficient $^{15}$N fractionation.

**Table 3**. Review of reactions involved in $^{15}$N fractionation.

$k_f = \alpha \times (T/300)^\beta \times f(B,m)/(f(B,m) + \exp(\Delta E/kT))$,
$k_r = \alpha \times (T/300)^\beta \times \exp(\Delta E/kT)/(f(B,m) + \exp(\Delta E/kT))$

| | Reaction | | α | β | ΔE(K) | f(B,m) | reference |
|---|---|---|---|---|---|---|---|
| 1. | $^{15}$N$^+$ + H$_2$ | → $^{15}$NH$^+$ + H | 4.2e-10 | 0 | 37.1 | - | (Roueff et al. 2015, Marquette *et al.* 1988) |
| 2. | H + $^{15}$NNH$^+$ | → N$^{15}$NH$^+$ + H | "0" | | -8.1 | | Barrier of 1500 K at M06-2X/VTZ level (this work, see appendix) |
| 3. | $^{15}$N + NNH$^+$ | → $^{15}$NNH$^+$ + N | "0" | | | | Large barrier, see Roueff et al. (2015) |
| 4. | $^{15}$N + NNH$^+$ | → N$^{15}$NH$^+$ + N | "0" | | | | Large barrier, see Roueff et al. (2015) |
| 5. | $^{15}$N + HCNH$^+$ | → HC$^{15}$NH$^+$ + N | "0" | | | | Large barrier, see Roueff et al. (2015) |
| 6. | $^{15}$N + CN | → C + $^{15}$NN<br>→ N + C$^{15}$N | 8.8e-11<br>0 | 0.42 | - | - | (Daranlot et al. 2012)<br>See text |
| 7. | $^{15}$N$^+$ + N$_2$ | → N$^+$ + N$^{15}$N | 3.3e-10 | 0 | -28.3 | 2 | (Fehsenfeld *et al.* 1974, Anicich *et al.* 1977) |
| 8. | N$^{15}$N + N$_2$H$^+$ | → N$^{15}$NH$^+$ + N$_2$ | 4.6e-10 | 0 | -10.3 | 0.5 | (Adams & Smith 1981) |
| 9. | N$^{15}$N + N$_2$H$^+$ | → $^{15}$NNH$^+$ + N$_2$ | 2.3e-10 | 0 | -2.1 | 0.5 | (Adams & Smith 1981) |
| 10. | N$^{15}$N + $^{15}$NNH$^+$ | → N$^{15}$NH$^+$ + N$^{15}$N | 2.3e-10 | 0 | -8.1 | 1 | (Adams & Smith 1981) |
| 11. | NH$_4^+$ + $^{15}$NH$_3$ | → $^{15}$NH$_4^+$ + NH$_3$ | 1.3e-9 | -0.5 | -14.5 | 1 | This work (capture rate constant assuming no barrier for proton exchange) |
| 12. | HCNH$^+$ + HC$^{15}$N | → HC$^{15}$NH$^+$ + HCN | 2.0e-9 | -0.5 | -10.1 | 1 | This work based on (Cotton et al. 2013) |
| 13. | HCNH$^+$ + H$^{15}$NC | → HC$^{15}$NH$^+$ + HCN<br>→ HCNH$^+$ + HC$^{15}$N | 1.0e-9<br>1.0e-9 | -0.5<br>-0.5 | 0<br>0 | | This work based on (Cotton et al. 2013) |
| 14. | HCNH$^+$ + e$^-$ | → HCN + H<br>→ HNC + H<br>→ CN + H + H | 9.62e-8<br>9.62e-8<br>9.06e-8 | -0.65<br>-0.65<br>-0.65 | 0<br>0<br>0 | -<br>-<br>- | Rate constant from (Semaniak *et al.* 2001)<br>Branching ratios from (Mendes *et al.* 2012, Barger *et al.* 2003, Herbst *et al.* 2000) |
| 15. | HC$^{15}$NH$^+$ + e$^-$ | → HC$^{15}$N + H<br>→ H$^{15}$NC + H<br>→ C$^{15}$N + H + H | 9.34e-8<br>9.90e-8<br>9.06e-8 | -0.65<br>-0.65<br>-0.65 | 0<br>0<br>0 | -<br>-<br>- | This work (see text) |

On grains, exchange reactions are not efficient in our model because we consider that the addition or bimolecular channels dominate, for example s-$^{15}$N + s-CN → s-$^{15}$NCN, s-CN$^{15}$N but not s-N + s-C$^{15}$N. As diffusion and tunneling are mass dependent they are not strictly equivalent for the various isotopologues. These effects are included in the model but they only have a small effect on $^{15}$N fractionation.

2.2 HCNH$^+$ + electron

Electronic Dissociative Recombination (DR) of HCNH$^+$ is a very important pathway for CN, HCN and HNC production. The experiments of (Mendes et al. 2012) showed that the majority of the exothermic energy released by the HCNH$^+$ + e$^-$ reaction is carried away as internal energy of the HCN-HNC products. This is in good agreement with the fact that about

30% of the HCN-HNC produced dissociates into CN + H. In interstellar clouds, the only way to relax this internal energy is through radiative emission of an infrared photon. As noted by Herbst et al. (2000), the typical time-scales for HNC-HCN inter-conversion is much shorter than relaxation by one infrared photon. Consequently, as relaxation occurs slowly, isomerization leads to balanced isomeric abundances at each internal energy. The final balance is determined at or near the effective barrier to isomerization, which corresponds to the energy of the transition state. The ratio between the isomeric forms is then approximated by the ratio of the ro-vibrational densities of states of the isomers at the isomerization barrier. Statistical theory leads to a HCN/HNC ratio close to 1 for the main isotopologues (Herbst et al. 2000). This result has been confirmed by Barger et al. (2003) through ab initio calculations of the radiative relaxation of HCN-HNC at energies above the barrier. Assuming a branching ratio of 1 for the main isotopologue production HCN/ HNC, statistical theory leads to a branching ratio equal to 0.91 for $HC^{15}N/ H^{15}NC$ considering the variation in the ro-vibrational densities of states, the branching ratio for CN production being the same for all isotopologues.

## 3 RESULTS

The time dependent abundances of the gas phase species $N_2H^+$, $NH_3$, s-$NH_3$, HCN, HNC, CN, $HC_3N$, $CH_3CN$ and $H_2CN$ (relative to $H_2$) calculated by our model are shown in Figure 1.

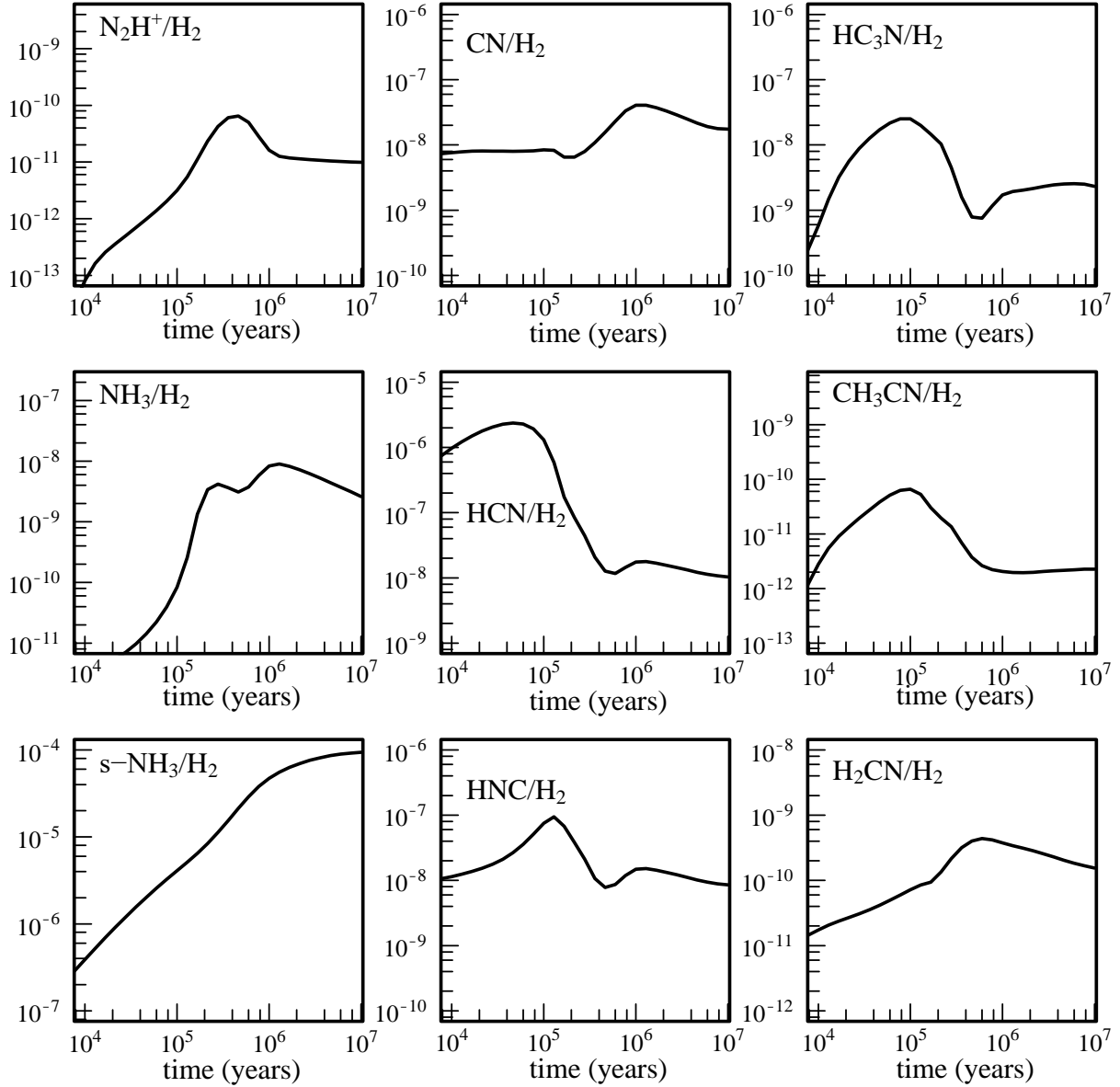

**Figure 1:** Gas phase species abundances (relative to $H_2$) of $N_2H^+$, $NH_3$, s-$NH_3$, HCN, HNC, CN, $HC_3N$, $CH_3CN$ and $H_2CN$ studied in this work as a function of time predicted by our model for $N(H_2) = 2\times10^4$ cm$^{-3}$ and T = 10K. The horizontal grey rectangles represent the abundances observed for TMC-1 (Crutcher *et al.* 1984, Ohishi *et al.* 1994, Pratap *et al.* 1997, Dickens *et al.* 1997, Turner *et al.* 1999, Gratier *et al.* 2016) and (Boogert *et al.* 2015) for s-$NH_3$ on Ice (Light Young Stellar Objects, LYSOs), including uncertainties. The vertical grey rectangles represent the TMC-1 cloud age given by the best agreement between the abundances given by our model and the observations ($N_2H^+$, $NH_3$, HCN, HNC, CN, $HC_3N$, $CH_3CN$ but also NO, CH, $C_2H$, c-$C_3H_2$, …)

Good agreement is obtained when our calculations are compared with the observed fractional abundances of the main isotopologues of HCN, HNC and CN (and many other species such as NO, CH, $C_2H$, c-$C_3H_2$, …) for typical dense molecular clouds with a cloud age around $(3-6)\times10^5$ years for a total density $n(H_2) = 2\times10^4$ cm$^3$ (the observed abundances shown in Figure 1 are those derived for TMC-1 molecular cloud (Crutcher et al. 1984, Ohishi et al. 1994, Pratap et al. 1997, Dickens et al. 1997, Turner et al. 1999, Gratier et al. 2016)).

The agreement is a little less good for $NH_3$, $N_2H^+$, $HC_3N$, $CH_3CN$ and $H_2CN$, reflecting our poorer understanding of nitrogen chemistry in dense molecular clouds. The underestimation of $NH_3$, $HC_3N$ and $CH_3CN$, with regard to previous versions of our model (Loison *et al.* 2014b, Loison *et al.* 2014a), are partly due to the introduction of the reactions with atomic carbon which have been recently measured for $C + NH_3$ (Hickson *et al.* 2015, Bourgalais *et al.* 2015) and calculated for $C + HC_3N$ (Li *et al.* 2006). The overestimation of $H_2CN$ may be due to an overestimation of the rate for the $N + CH_3$ reaction at 10 K, the main source of $H_2CN$ (Hébrard *et al.* 2012). The rate constant for this reaction could decrease at low temperature (Marston *et al.* 1989), so this process should be reinvestigated at even lower temperatures.

The time dependent $^{14}N/^{15}N$ ratios calculated by our model for the main nitrogen compounds in the gas phase are shown in Figure 2.

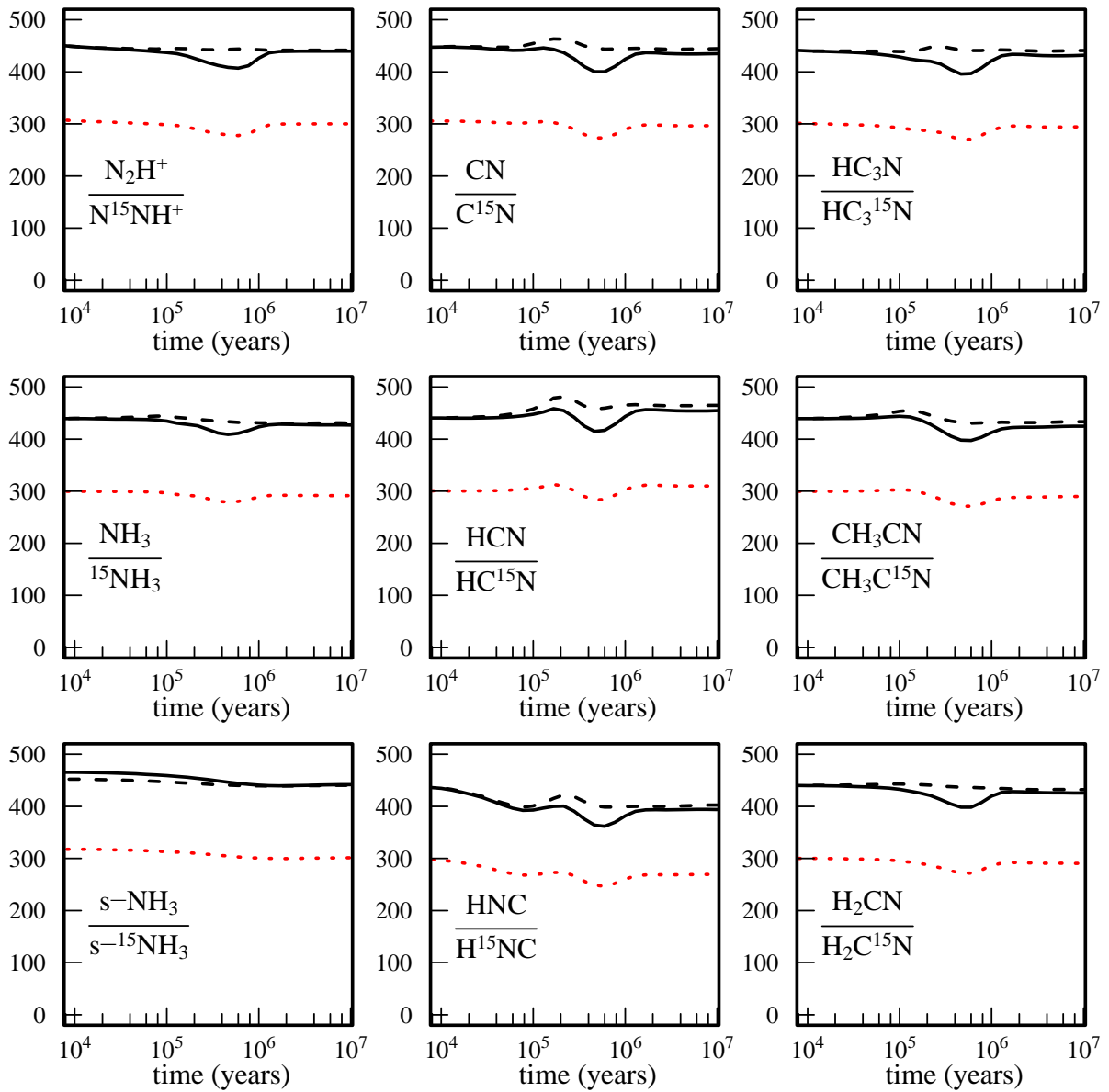

**Figure 2:** Calculated N/$^{15}$N ratio for the main nitrogen species in the gas phase as a function of time predicted by our model (N(H$_2$) = 2×10$^4$ cm$^{-3}$, T = 10K) (for an elemental N/$^{15}$N ratio equal to 441 in continuous black lines, and equal to 300 in red dotted lines). With our model both N$^{15}$NH$^+$ and $^{15}$NNH$^+$ have very similar $^{15}$N fractionation levels. In black dashed lines: model with the same sticking rate constant for all isotopologues of a given species (same rate constant for N$_2$ and N$^{15}$N, …) for an elemental N/$^{15}$N ratio equal to 441. The vertical grey rectangles represent the TMC-1 cloud age as described in Figure 1.

Our model leads to low nitrogen fractionation levels. This result demonstrates that neither gas-phase isotopic exchange reactions (see Table 3) nor diffusion on grains have significant effects on nitrogen fractionation in dense interstellar clouds. Although the overall fractionation levels are low, certain results should be highlighted. First, the lower $^{14}$N/$^{15}$N ratios for HNC derive from the dissociative recombination of HCNH$^+$, a reaction leading to more $^{15}$N enrichment for HNC than for HCN and CN as explained in the model description section. Another nitrogen fractionation pathway occurs through mass dependent accretion. Here, the accretion time for each isotopologue of a given species is dependent on its mass, so that in the case of atomic nitrogen, $^{14}$N will be removed slightly more quickly from the gas phase than $^{15}$N. Assuming an initial elemental $^{14}$N/$^{15}$N ratio in the local ISM equal to 441, for a typical cloud age around 3-6×10$^5$ years and a total density of N(H$_2$) = 2×10$^4$ cm$^{-3}$, the model leads to some nitrogen fractionation, with $^{14}$N/$^{15}$N ratios around 400 for most of the species (N$_2$H$^+$, NH$_3$, HCN, CN, …) and 360 for HNC. This depletion induced fractionation effect for nitrogen is only efficient because atomic nitrogen is relatively unreactive in the gas-phase (Daranlot *et al.* 2013, Daranlot et al. 2012, Daranlot *et al.* 2011), so that sticking onto grains is the dominant loss process. Although the difference of the accretion rate between $^{14}$N and $^{15}$N is small, the effect is large enough to remove a few tens of percent more $^{14}$N atoms from the gas-phase when the accretion time reaches 4 × 10$^5$ years. As the remaining atomic nitrogen in the gas-phase is depleted in $^{14}$N, all nitrogen bearing species formed through gas-phase chemistry will reflect these higher $^{15}$N levels with various values depending on the individual formation mechanisms (such as N + CH → CN + H for example). Although the calculated $^{14}$N/$^{15}$N values for NH$_3$, HCN, HNC, CN and HC$_3$N are lower than those obtained in earlier models they are still above the observed values shown in Table 1. The effect is not very large for evolved clouds because for cloud chemical ages in the range 3-6×10$^5$ years, most of the nitrogen has already been removed from the gas-phase. So even if the gas-phase is depleted in $^{14}$N, the $^{14}$N/$^{15}$N fraction for species formed on grains is close to or slightly above the elemental $^{14}$N/$^{15}$N ratio in the local ISM. Desorption mechanisms will release $^{14}$N-bearing molecules back into the gas-phase resulting in a reduction in the gas-phase fractionation levels for these species.

Besides the uncertainties generated by the desorption mechanisms, grain depletion is the most efficient fractionation process in our model. This leads to lower $^{14}N/^{15}N$ ratios for species formed in the gas-phase in dense molecular clouds (CN, HCN, HNC) than the solar wind reference value (441 ± 2.5 (Marty et al. 2010) corresponding to the nitrogen reservoir of the prestellar cloud). However, mass dependent sticking rate cannot explain $^{14}N/^{15}N$ ratios as low as those listed in Table 1, except when larger uncertainties than those reported in the literature are considered. Nevertheless, considering a lower initial elemental $^{14}N/^{15}N$ ratio in the local ISM, around 300 from (Romano *et al.* 2017), the inclusion of mass dependent accretion for N atoms results in agreement with most observations for CN, HCN, HNC, $HC_3N$ and $NH_3$ (except HCN and HNC in B1 and $HC_3N$ in L1544, values shown in Table 1), the average for all these observations being equal to 312 ± 48, but also for $HC_5N$ (Taniguchi & Saito 2017).

The comparison with previous studies is not obvious due to the large differences between models. The low nitrogen fractionation induced by chemistry is similar in Roueff et al. (2015) despite the large differences in models (no grain chemistry in Roueff et al. (2015) and substantial differences in the chemical networks). The low fractionation is due in both cases to the absence of efficient fractionation reactions. Wirström and Charnley (2018) found some nitrogen fractionation in their recent work. In their model, $NH_3$ in the gas phase shows some $^{15}N$ depletion due to specific reactions of $^{15}N^+$ in conjunction with a large gas phase $N_2$ abundance. The large $N_2$ abundance is due to the fact that Wirström and Charnley (2018) consider that N and $N_2$ do not freeze out, which is not the case in our model. They also found some $^{15}N$ enrichment in CN and HCN due to the $^{15}N$ + CN reaction. However, this reaction is likely to be inefficient (see section 2.1). Moreover, as N atom grain depletion is not considered by these authors, the $^{14}N/^{15}N$ mass dependent accretion rate cannot be reproduced. In their recent study, Furuya and Aikawa (2018) developed a full gas-grain model for $^{15}N$ fractionation including the first step of dense molecular cloud formation where UV photons can penetrate. They obtained some $^{15}N$ fractionation due essentially to the isotope selective photodissociation of $N_2$, whereas chemistry is inefficient. As we consider $A_v = 10$ in our study we cannot reproduce such effects. In their work, they consider the difference in the adsorption rates for $^{14}N$ and $^{15}N$. However, in their model this effect is likely to be significantly smaller than the $^{15}N$ atomic enrichment in the gas phase due to the photodissociation of $N^{15}N$. This is because N is depleted onto grains when $A_v$ is in the 0-2 range where isotope selective photodissociation of $N_2$ is at work and very efficient.

## 3.1 $^{15}NNH^+$, $N^{15}NH^+$ fractionation

Our model produces similar low $^{15}N$ fractionation effects for N-bearing species formed by gas-phase reactions. Simulated isotopologue abundances for $N_2H^+$ show that fractionation levels are small and incompatible with observations in L1544 (Redaelli et al. 2018). Little enrichment occurs for $N_2H^+$ in our model as the exchange reaction $N^{15}N + N_2H^+ \rightarrow N^{15}NH^+/^{15}NNH^+ + N_2$ is inefficient due to its low exothermicity. In addition, the DR reaction of $N_2H^+$ and its reaction with CO are much more efficient sinks of $N_2H^+$. It should be noted that with our model, both $N^{15}NH^+$ and $^{15}NNH^+$ have very similar low $^{15}N$ fractionation levels. Specifically, the $N^{15}N + NNH^+ \rightarrow NN + N^{15}NH^+/^{15}NNH^+$ reactions are inefficient despite large rate constant at 10K because the flux of these reactions is negligibly small compared to the fluxes of DR and the reaction with CO. The chemistry of $N_2H^+$ is essentially driven by three main reactions:

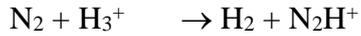
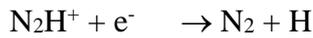
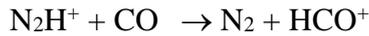

As the $N_2 + H_3^+$ and $N_2H^+ + CO$ reactions are direct barrierless processes, they should not be affected by the zero-point energy (ZPE) difference of any of the $N_2H^+$ isotopologues. In our model, we assume that the rate constants for DR of $^{15}NNH^+$, $N^{15}NH^+$ and $N_2H^+$ are identical. However, Lawson *et al.* (2011) showed that the DR of $N_2H^+$, $^{15}N_2H^+$ and $N_2D^+$ have notable differences (more than 20%) between 300 K and 500 K, the DR reaction of $N_2H^+$ being more efficient that the DR reactions of $^{15}N_2H^+$ and $N_2D^+$. However, that result does not necessarily mean that the DR reactions of $N^{15}NH^+$ and $^{15}NNH^+$ are less efficient than $N_2H^+$ at low temperature due to their complexity. Indeed, the difference in ZPE may favor intercrossing curves between the ions and the neutral for $^{15}NNH^+$ and $N^{15}NH^+$ leading to an unexpected increase at low temperature. Experimental measurements of the DR rates of $^{15}NNH^+$ and $N^{15}NH^+$ are clearly needed but higher DR rate constants for $N^{15}NH^+$ and $^{15}NNH^+$ than the equivalent $N_2H^+$ one could be possible. If the DR reactions of $N^{15}NH^+$ and $N^{15}NH^+$ are more efficient than the $N_2H^+$ one, the $N_2H^+/N^{15}NH^+$ and $N_2H^+/N^{15}NH^+$ ratios will increase above and beyond the input nitrogen isotope ratio value. When gas-phase CO is low, the main loss for $N_2H^+$ is DR. Then, if the rate constants for the DR of $^{15}NNH^+$ and $N^{15}NH^+$ are more efficient than the DR of $N_2H^+$, the $N_2H^+/^{15}NNH^+$ and $N_2H^+/N^{15}NH^+$ ratios are notably larger than the elemental $N/^{15}N$ ratio, around 900 for the conditions used in Figure 3. However, when the CO abundance is large, the $N_2H^+ + CO$ reaction is competitive with DR and the

$N_2H^+/^{15}NNH^+$ and $N_2H^+/N^{15}NH^+$ ratios are closer to the elemental $N/^{15}N$ ratio, namely around 620 for the conditions used in Figure 3. It should be noted that this effect is highly dependent on the rate constants of the DR reactions of $N_2H^+$, $^{15}NNH^+$ and $N^{15}NH^+$ in addition to the value of the rate constant for the $N_2H^+$ + CO reaction measured only once (Bohme *et al.* 1980) as well as the total density of the molecular cloud.

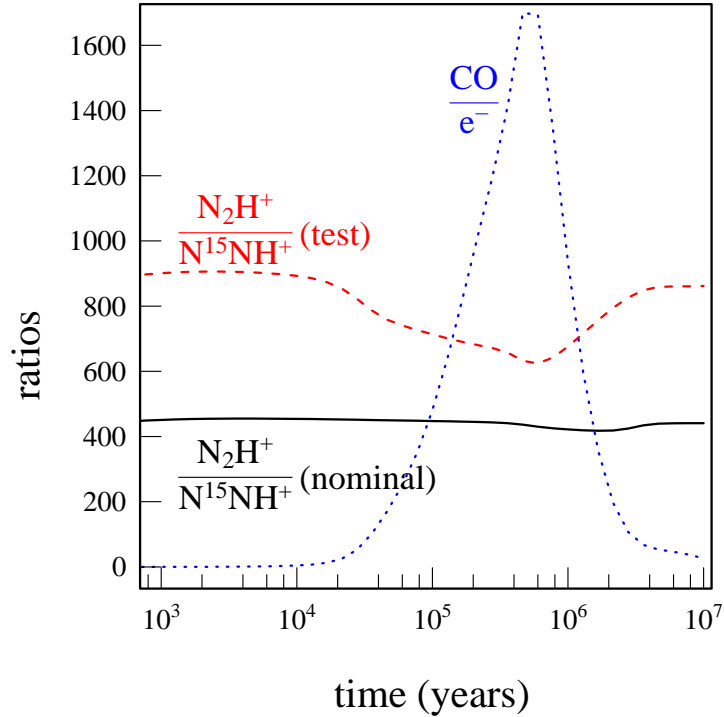

**Figure 3:** Calculated $N_2H^+/N^{15}NH^+$ ratios for the nominal model (continuous black line) and considering a $N_2H^+$ DR rate constant divided by two (test, dashed red line), $n(H_2) = 2\times10^4$ cm$^{-3}$, T = 10K. The CO/e$^-$ ratio (dotted blue lines) is also plotted showing that when CO abundance reaches its maximum, the CO + $N_2H^+$ reaction become competitive with the $N_2H^+$ + e$^-$ reaction.

An interesting observation that supports this hypothesis is the evolution of the abundance ratios $N_2H^+/N^{15}NH^+$, $N_2H^+/N^{15}NH^+$ and $N_2H^+/N_2D^+$ toward various high mass star-forming cores (Fontani *et al.* 2015). In these regions, the $N_2H^+/N^{15}NH^+$ and $N_2H^+/N^{15}NH^+$ ratios are seen to increase with the mass of the core, while the corresponding $N_2H^+/N_2D^+$ ratio, whose value is known to be a good indicator of the level of CO depletion (Roberts *et al.* 2003, Crapsi *et al.* 2005), is seen to decrease for the same objects.

4 CONCLUSION

Despite a thorough search initiated in (Roueff et al. 2015) and continued in this work, chemistry does not appear to lead to significant nitrogen fractionation. The most efficient fraction processes are the mass dependent fractionation mechanism induced by depletion onto grains and DR of HCNH$^+$, although these effects produce relatively low fractionation levels. This low nitrogen fractionation may not be incompatible with observations. Considering the uncertainties brought about by saturation of the main isotopologues transitions, nitrogen fractionation in dense molecular clouds may not be so different to the "elemental" $^{14}$N/$^{15}$N ratio in the local ISM, of approximately 300 (Adande & Ziurys 2012). Indeed, apart from N$_2$H$^+$, all other nitrogen compounds (CN, HCN, HNC, HC$_3$N and NH$_3$, described in Table 1, and HC$_5$N (Taniguchi & Saito 2017)) seem to show similar nitrogen fractionation levels in dense molecular clouds, namely 312 ± 48 from values in Table 1 (without all N$_2$H$^+$ observations nor those for HCN and HNC in B1). The difference between the "elemental" $^{14}$N/$^{15}$N ratio in the local ISM and the solar wind reference value, 441 ± 5 (Marty et al. 2010), being in that case explained through galactic chemical evolution (Romano et al. 2017). However, this scenario cannot explain the observed HC$_5$N fractionation in L1544 (400 ± 20) (Hily-Blant et al. 2018)), neither can it explain the strong observed enrichment in HCN and HNC in B1 (Daniel et al. 2013). Additionally, different dissociative recombination rates are required to explain the observed N$_2$H$^+$ fractionation levels, which need experimental confirmation. Moreover, Colzi et al. (2018) recently found a $^{14}$N/$^{15}$N ratio for HCN and HNC across the galaxy much closer to 400, which challenges the currently adopted $^{14}$N/$^{15}$N ratio value of 300 in the local ISM.

An alternative scenario to explain nitrogen fractionation in dense molecular clouds involves N$_2$ photodissociation. Indeed, N$_2$ photodissociation is the only efficient process leading to significant $^{15}$N enrichment in a wide range of astrochemical environments such as protoplanetary discs, comets and planetary atmospheres (Liang *et al.* 2007, Li *et al.* 2013, Heays et al. 2014, Dobrijevic & Loison 2018). However, to be efficient, isotope selective N$_2$ photodissociation requires a large N$_2$ column density in conjunction with low A$_v$. Furuya and Aikawa (2018) developed a model where N$_2$ and N$^{15}$N photodissociation in translucent molecular clouds (with low visual extinction) at the outset of molecular cloud formation can induce nitrogen fractionation. In the translucent part of the molecular clouds, N$_2$ becomes $^{15}$N depleted while CN, HCN, HNC and NH$_3$ become $^{15}$N enriched. The N$^{15}$N depletion is partly conserved in dense molecular cloud but their model cannot reproduce the high $^{15}$N depletion of N$_2$H$^+$ observed in L1544 and L429 (Redaelli et al. 2018). Moreover, the $^{15}$N enrichment is not conserved in the gas phase of dense molecular clouds because N atoms are efficiently transformed into s-NH$_3$ on ices, which becomes the main $^{15}$N reservoir. As s-NH$_3$ is not easily desorbed, most of the gas phase nitrogen chemistry in the dense molecular clouds is initiated by N$_2$ dissociation (through cosmic rays or through its reaction with He$^+$). As N$_2$ is depleted in

$^{15}$N, the chemistry induced by N$_2$ dissociation in the dense clouds counterbalance the enrichment due to N$_2$ photodissociation in the translucent cloud. Then the Furuya and Aikawa (2018) model requires low elemental $^{15}$N fractionation levels in the local ISM to reproduce the observations. The s-NH$_3$ trapping effect is very efficient in the Furuya and Aikawa (2018) work because NH$_3$ photodesorption is inefficient and chemical desorption was fixed to 1% following Garrod et al. (2007). However, this amount can be much higher for addition to atoms and small species as shown by Minissale *et al.* (2016). Then the rapid hydrogenation of s-N, s-NH and s-NH$_2$ may lead to much higher NH, NH$_2$ and NH$_3$ abundances in the gas phase. Additionally, the photodissociation of s-NH$_3$ followed by the s-H + s-NH$_2$ and s-H + s-NH reactions (NH + H + H are the favored products for NH$_3$ photodissociation at 121.6 nm (Slanger & Black 1982)), may strongly enhance chemical desorption through cycling processes. Then, s-NH$_3$ may not be such an important reservoir and the release of NH$_2$ and NH$_3$ in the gas phase (with NH$_2$ and NH$_3$ enriched in $^{15}$N) may induce an efficient nitrogen chemistry, potentially much more efficient that the one induced by than N$_2$ dissociation. A particularly interesting aspect of the Furuya and Aikawa (2018) model is that it may explain variable $^{15}$N fractionation levels from cloud to cloud depending on the formation history.

Our model clearly demonstrates that chemistry cannot lead to high nitrogen fractionation levels but there remains much work to be done to clarify the nitrogen fractionation issue from diffuse clouds to the solar system. Among these issues, an in-depth review of the various observations is required to assess the dispersion of nitrogen fractionation data. Also, dissociative recombination rates for the various N$_2$H$^+$ isotopologues need to be measured. Finally, a full molecular cloud formation model such as the one developed by Furuya and Aikawa (2018) seems to be an indispensable approach for the future, requiring however better characterization of critical processes such as chemical desorption.

This work was supported by the program "Physique et Chimie du Milieu Interstellaire" (PCMI) funded by CNRS and CNES. VW researches are funded by the ERC Starting Grant (3DICE, grant agreement 336474). Computer time was provided by the Pôle Modélisation HPC facilities of the Institut des Sciences Moléculaires UMR 5255 CNRS − Université de Bordeaux, co-funded by the Nouvelle Aquitaine region.

**Appendix: Theoretical calculations on the H + $^{15}$NNH$^+$ → H + N$^{15}$NH$^+$ reaction**

Relative energies at the M06-2X/VTZ (in kJ/mol at 0 K including ZPE) with respect to the H + N$_2$H$^+$ energy, geometries and frequencies (in cm$^{-1}$, unscaled) of the various stationary points. The absolute energies at the M06-2X/VTZ level including ZPE in hartree are also given in column 1.

| Species (Energy, hartree) | Relative energies (kJ/mol) | Geometries (Angstrom) | Harmonic Frequencies (cm$^{-1}$) |
|---|---|---|---|
| H (-0.4981348) | | | |
| $^{15}$NNH$^+$ (-109.708017) 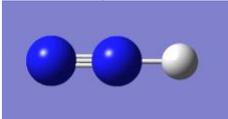 | 0 | N  0.0  0.0   0.646077<br>N  0.0  0.0  -0.435410<br>N  0.0  0.0  -1.474664 | 755, 755, 2406, 3395 |
| TS (-110.201440) 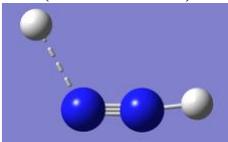 | +12.4 | N  -0.106902  -0.555629  0.0<br>N  -0.106902   0.539452  0.0<br>H   0.018049   1.565755  0.0<br>H   1.478578  -1.452517  0.0 | 459, 545, 760, 2264, 3390,<br>814i (imaginary frequency) |